\newcommand{\Pl}{\partial}
\newcommand{\ts}{\textstyle}
\newcommand{\fder}[2]{\frac{{\ts d \/ #1}}{{\ts d\/ #2}}}
\newcommand{\fpar}[2]{\frac{{\ts \Pl \/ #1}}{{\ts \Pl \/ #2}}}
\newcommand{\nder}[3]{\frac{{\ts d^{#1} \/ #2}}{{\ts d \/ #3^{#1}}}}

\newcommand{\bee}{\begin{equation}}
\newcommand{\ene}{\end{equation}}
\newcommand{\beea}{\begin{eqnarray}}
\newcommand{\enea}{\end{eqnarray}}

\documentclass[prb,aps,twocolumn]{revtex4}

\usepackage{dcolumn}
\usepackage{graphicx}
\usepackage{latexsym}
\usepackage{amsmath}
\usepackage{amssymb}
\usepackage{subfigure}
\usepackage{soul}

\begin{document}
 
\title{Excitation of Electrostatic Standing Wave in the Superposition of Two Counter Propagating Relativistic Whistler Waves}
\author{Mithun Karmakar$^{1}$, Sudip Sengupta$^{1,2}$ and Bhavesh Patel$^{1}$ }
\affiliation{$^1$Institute for Plasma Research, Bhat, Gandhinagar 382428, India\\
$^2$Homi Bhabha National Institute, Training School Complex, Anushakti Nagar, Mumbai 400085, India}

\begin{abstract}
The problem of standing wave formation by superposing two counter-propagating whistler waves {\it in an overdense plasma}, studied recently by Sano {\it et al.} (Phys. Rev. E {\bf 100}, 053205 (2019) and Phys. Rev. E {\bf 101}, 013206  (2020)), has been revisited in the relativistic limit. A detailed theory along with simulation has been performed to study the standing wave formation in the interaction of two counter propagating relativistically intense whistler waves. The relativistic theory explains such interaction process  more precisely and predicts correct field amplitudes of the standing wave for a much wider range of physical parameters of the problem as compared to its non-relativistic counterpart.
%
%
The analytical results are compared with 1-D Particle-in-Cell (PIC) simulation results, performed using OSIRIS 4.0.  
The results are of relevance to ion heating and fast ignition scheme of inertial confinement fusion.
\end{abstract}

\maketitle
 
\begin{section}{Introduction}
Acceleration and heating of plasma ions by using large amplitude electromagnetic waves or ultra-intense laser pulses is a very useful and efficient method to directly couple the electromagnetic field energy to ions and thermalize them.\cite{roth,naumova,lindl,atzeni,sano,sano1} The basic idea of one such ion heating mechanism associated with whistler waves has been recently examined by Sano {\it et al.}. \cite{sano,sano1} 
They have investigated the problem of ion heating in the interaction of whistler waves analytically by adopting a simple theoretical model and also numerically by performing  Particle-in-cell simulation. Whistler waves have some important and essential characteristic features which make this electromagnetic plasma wave mode a suitable candidate for this purpose.\citep{akheizer,chen}   In this ion acceleration and heating mechanism, initially a standing wave is formed by the superposition of two counter propagating whistler waves. The ions are  accelerated by the electrostatic field of the standing wave, resulting in the excitation of an unstable ion wave through streaming instabilities\cite{chen,rupen1,rupen2}, which eventually heats the ions through collapse of the ion wave itself. Ion heating by whistler waves has importance in many research field starting from laboratory plasmas to space and astrophysical plasmas, in the diagnostics of the material structures, medicine, security and industrial engineering.\cite{huba,perez} Also in plasma-based fusion science, it is necessary to heat the plasma ions for efficient fusion reaction to occur.\cite{roth,naumova,lindl,atzeni}


Whistler wave is basically a low frequency right circularly polarized (RCP) electromagnetic mode having a frequency which is lower than the electron cyclotron frequency associated with the ambient constant magnetic field. Such a mode is excited in a plasma system when an external magnetic field with sufficient amplitude is applied along 
the propagation direction. The characteristic features and propagation characteristics of whistler waves have been studied extensively in the past.\cite{sano,sano1,akheizer,chen,huba,perez,luan}
This mode has a unique attribute of non-existence of cut-off which enables it to propagate through over-dense plasma system\cite{sano,yang,ma,weng,clemmow}.
The relativistic theory developed by Akheizer and Polovin in 1956 \cite{akheizer} clearly shows how even in over-dense plasma system whistler waves can readily propagate without any cut-off density.
In a recent investigation the transmission and propagation properties of whistler waves in over dense plasma system have been studied by Luan {\it et al}.\cite{luan} using Particle-in-cell simulation. 
In another investigation the interaction of relativistic whistler waves with dense plasma has been performed.\cite{lv}  The impact of external axial magnetic field on the proton acceleration has been discussed when the target plasma is irradiated by the whistler modes. It has been shown that proton can be accelerated to around 15 MeV energy with an external magnetic field of the order of 20-25 kT with femtosecond laser of intensity around 5$\times$ 10$^{18}$ W/cm$^2$.

Our present investigation is primarily motivated by the work of Sano et al.\cite{sano,sano1}. The non-relativistic theory which they have developed for the excitation of standing waves by superposition of whistler waves, in single ion species plasma, reveals beautifully the interesting physical processes associated with it and demonstrates the ion heating process. It has been shown that ion can be heated up to several keV by the collapse of the standing wave, for their chosen system parameters. They have further extended their analysis by choosing a multi-ion species target like deuterium-tritium (DT) ices and solid ammonia borane ($H_6BN$) and reported similar findings.\cite{sano1}. However, a complete and general theory of  high intensity laser matter interaction process must take relativistic effects into account. For ultra intense laser pulses, the 
relativistic effects can significantly change the propagation characteristics of whistler waves and also the energy exchange process with background ions. The relativistic theory enables us to correctly estimate the field amplitude of the standing wave for a much  wider range of physical parameters (laser intensity and external magnetic field) as compared to its non-relativistic counterpart.

In this paper we have attempted to establish a relativistic theory of superposition of whistler waves and consequent standing wave 
formation. In contrast to the non-relativistic theory, the refractive index of the plasma associated with relativistically intense whistler waves is not a constant, but it varies with the relativistic Lorentz factor associated with the electron fluid velocity\cite{akheizer}. 
This sole feature differentiates this present investigation on standing wave formation from its non-relativistic counterpart and brings out the new physics associated with it. One essential outcome of the relativistic theory is that, it restricts the possible values of the external magnetic field which are needed to excite the standing wave, for a given laser intensity. It turns out that for intensities of the order $\sim 10^{21} W/cm^{2}$ or more, the required magnetic field becomes several times higher than that allowed by non-relativistic theory.
Existence of such  strong external magnetic field helps to greatly enhance the conversion efficiency of the laser energy to the plasma ions or electrons\cite{cai,johzaki}.
In astrophysical objects like pulsar and magnetars, very high magnetic field of the order of giga gauss to tera gauss can be observed.\cite{archi,turo,beskin} Experimentally, a highest magnetic field of 70 kT has been reported 
so far in interaction of high intensity laser ( $9 \times 10^{19}$ W/cm$^2$ ) with a plasma of density 10$^{19}$- 10$^{20}$ cm$^{-3}$. \cite{wagner}
So, our present investigation is quite timely and it has its relevance to laboratory laser matter interaction as well as in the astrophysical situations.

The paper is organized as follows. In section II, we present the  dispersion relation for relativistic whistler waves by following the theory developed by Akheizer and Polovin\cite{akheizer}. This is presented here for the sake of completeness. In third section, the relativistic theory is developed to estimate the amplitude of the electric field of the standing wave which is excited in the superposition of counter propagating whistler waves. The results of our investigation in the non-relativistic limit is also discussed and compared with the earlier works. In section IV, we present 1-D PIC simulation results obtained using OSIRIS 4.0 and a comparison with our theoretical findings. Finally we summarize our work 
in section V.

\end{section}


\section{Basic Formulation}
The derivation presented in this section has been described
earlier by Akheizher and Polovin.\cite{akheizer} As stated in the introduction  here we present it for the sake of  completeness and to get a clear idea of the problem. We start with the basic set of Maxwell's equations along with the relativistic fluid equations to arrive at the relativistic dispersion relation for the whistler waves. The full set of fluid-Maxwell's equations in a cold plasma, in three-dimensions can be written as,
\beea
\nabla \times {\bf E} =-\frac{1}{c}\fpar{{\bf B}}{t},
\label{an1}
\enea
\beea
\nabla \times {\bf B} =\frac{1}{c}\fpar{{\bf E}}{t}-\frac{4\pi}{c}n_e e{\bf v},
\label{an2}
\enea
\beea
\nabla \cdotp {\bf E} =4 \pi e(n_0-n_e),
\label{an3}
\enea
\beea
\nabla \cdotp {\bf B} =0,
\label{an4}
\enea
\beea
\fpar{n_e}{t}+ \nabla \cdotp (n_e {\bf v})=0.
\label{an6}
\enea
\beea
\fpar{{\bf p}}{t}+({\bf v}\cdotp \nabla){\bf p}=-e {\bf E}-
\frac{e}{c}({\bf v }\times {\bf B}).
\label{an5}
\enea
Here, ${\bf p}=m{\bf v}/\sqrt{1-v^2/c^2}$, $n_0$ is the equilibrium density and $n_e$ is the electron density. Other variables have their usual meanings. Since, we are only interested in the relativistic whistler wave dynamics in one dimension, we consider a specialized geometry of our problem. Both the electromagnetic wave vector and the external magnetic field ${\bf B_0}$ are assumed to be directed along positive z-axis. We now look for stationary wave frame solution which is a function of $\tau=(t -z/V)$. In this transformed co-ordinate, equations (\ref{an1}) - (\ref{an5}) may be combined as 
\beea
\nder{2}{\rho_x}{\tau}+\beta \Omega\fder{}{\tau}\left(\frac{u_y}{\beta - u_z}\right)+\frac{\beta^2}{\beta^2-1}\frac{\beta}{\beta - u_z}\omega_p^2u_x=0
\label{n11}
\enea
\beea
\nder{2}{\rho_y}{\tau}-\beta \Omega\fder{}{\tau}\left(\frac{u_x}{\beta - u_z}\right)+\frac{\beta^2}{\beta^2-1}\frac{\beta}{\beta - u_z}\omega_p^2u_y=0,
\label{n12}
\enea

\beea
\nder{2}{}{\tau}\left({\beta \rho_z-\sqrt{1+\rho^2}}\right)+\frac{\beta^2}{\beta-u_z}\omega_p^2u_z=0
\label{n13}
\enea
where, $\vec{\rho} = \mathbf{p}/mc$, $\mathbf{u}=\mathbf{v}/c$, 
$\Omega=eB_0/mc$ and $\beta=V/c$ being the normalized phase velocity of the plasma wave. Further, the components of electromagnetic fields may be computed in terms of fluid momenta as
\beea
\tilde{\bf B} = - \frac{1}{\omega \beta}\left(\hat{z}\times \fder{\vec{\rho}}{\tau}\right)+\left(\frac{ \mathbf{\Omega^{\prime}} \beta - \mathbf{u} \Omega^{\prime}}{\beta - u_z}\right)
\label{n61}
\enea
and
\beea
\tilde{\bf B}=\frac{1}{\beta}\left(\hat{z} \times \tilde{\bf E}\right) + {\bf \Omega^{\prime}}
\label{n71}
\enea
where, $\tilde{\bf{B}}=e{\bf B}/m\omega c, \tilde{\bf{E}}=e{\bf E}/m \omega c$, and ${\bf {\Omega}}^{\prime}={\bf \Omega}/\omega = e{\bf B_{0}} /m \omega c$.

For purely transverse oscillation $u_z=0$, and if we require bounded solution then Eq.(\ref{n13}) gives  $\rho^2=$ constant.\cite{akheizer,kaw} This implies that $u_x^2+u_y^2=u^2=$ constant. 
We substitute
\beea
{\bf u}= u(\cos \omega \tau \hat{x}+\sin \omega \tau\hat{y})
\label{n3}
\enea
in one of the the momentum equations [Eq. (\ref{n11}) or (\ref{n12})] to obtain following relativisitic dispersion relation for a purely transverse mode \cite{akheizer,clemmow,chen}
\beea
\omega=\frac{\bar{\Omega}}{2}\pm \left[\frac{\bar{\Omega}^2}{4}+\frac{\beta^2 {\bar{\omega}_p}^2}{\beta^2-1}\right]^{1/2}
\label{n4}
\enea
where, $\bar{\Omega}=\Omega\sqrt{1-u^2}$ and $\bar{\omega}_p^2=\omega_p^2\sqrt{1-u^2}$ are respectively the relativistically correct cyclotron frequency and plasma frequency.
The dielectric constant of the plasma is then given by 
\beea
\varepsilon=\beta^{-2}=1-\frac{\bar{\omega}_p^2}{\omega^2-\omega\bar{\Omega}}.
\label{n51}
\enea
For $\omega < \bar{\Omega}$, this is the dispersion relation for the Whistler branch, which may also be written differently as 
\beea
\varepsilon=1-\frac{{{\omega}_p^{\prime}}^2}{\gamma-{\Omega}^{\prime}}.
\label{n5}
\enea
where, ${\omega}_p^{\prime}=\omega_p/\omega $, 
${\Omega}^{\prime}=\Omega/\omega$.
From this expression it is evident that when $\gamma < {\Omega}^{\prime}$ there is no cut-off for the whistler waves even in highly dense plasma system. 
%
%
Finally using equations (\ref{n61}) and (\ref{n71}), the magnitude of electric and magnetic field amplitude in terms of fluid velocity ( and momentum ) may be respectively written as
%
\beea
\tilde E = \rho \left({\Omega}^{\prime}\sqrt{1-u^2}-1\right)
\label{n9}
\enea
\beea
\tilde B = \frac{\tilde E}{\beta}= \frac{\rho}{\beta} \left({\Omega}^{\prime}\sqrt{1-u^2}-1\right)
\label{n91}
\enea
These fields are entirely transverse.


\section{Standing Wave Formation: Superposition of two counter propagating whistler waves }
It is quite evident from the analysis in the preceding section, that for a single whistler wave propagating through plasma, formation of  electrostatic standing wave is not possible. Whistler waves do not couple to longitudinal perturbations. If however, we launch two counter propagating whistler waves from two sides of a plasma slab, an electrostatic standing wave will form due to the superposition of these counter propagating waves. In this section we shall discuss elaborately the excitation mechanism of such standing wave in the superposition of two counter propagating relativistically intense whistler waves.
The right circularly polarized forward ($+$ve) and backward propagating ($-$ve) whistler modes in the presence
of an external constant magnetic field ${\bf B_0}$ directed along the positive z-axis, can be described as
\beea
{\bf \tilde E_{\pm}} &=& \tilde E_{\pm}\exp{i(\omega t \mp k z)}(i\hat{x}+\hat{y}),\\
{\bf \tilde B_{\pm}} &=& \mp \tilde B_{\pm}\exp{i(\omega t \mp k z)}(\hat{x}-i\hat{y}).
\label{n7}
\enea
The associated fluid velocities are 
\beea
{\bf u_{\pm}} = u_{\pm}\exp{i(\omega t \mp k z)}(\hat{x}-i\hat{y}).
\label{•}
\enea
These two counter propagating whistler waves superpose to form a standing wave pattern in the plasma system producing an electrostatic field ($\tilde E_z$) in the longitudinal direction. The expression for the electric field of this standing wave calculated from
z-component of the electron momentum equation is given by
\beea
 \tilde E_z&=&-[({\bf u_+}+{\bf u_-})\times ({\bf \tilde B_+}+{\bf \tilde B_-})]_z\nonumber\\
&=& -(u_+\tilde B_-+u_-\tilde B_+)\sin (2k_0 N z).
\label{n8}
\enea
where $k=\omega/V=(\omega/c)/(V/c)=k_0/\beta=k_0N$, $N$ being the refractive index. Here $\tilde{B}_{\pm}$ is related to $u_{\pm}$ through equation (\ref{n91}); thus the only remaining unknown in the above expression is $u_{\pm}$. This is evaluated by equating 
the amplitude of the transmitted whistler waves ($\tilde {E_{\pm}}$)  to the fields inside the plasma as given by equation (\ref{n9}). The transmitted fields are given by Fresnel equations\cite{grifith} as   $\tilde E_{\pm}=2 a_0 / (N+1)$ where  ($a_0=eE_0/m \omega c$ )  is the free space laser amplitude.
Here N, the refractive index of the plasma, which can be obtained from the relativistic dispersion relation derived in the previous section, is further related to $u_{\pm}$ as
\beea
N=\beta^{-1}=\left[1-\frac{{{\omega}_p^{\prime}}^2\sqrt{1-u_{\pm}^2}}{1-{\Omega}^{\prime}\sqrt{1-u_{\pm}^2}}\right]^{1/2}, 
\label{n10}
\enea
where ${\omega}_p^{\prime}=\omega_p/\omega$ and ${\Omega}^{\prime}={ \Omega}/\omega$. This is important, as it differentiates the relativistic theory from its non-relativistic version. Putting all the components together, leads to an algebraic equation for $u_{\pm}$ as $f(x) = 0$, where f(x) is given by 
\begin{eqnarray}
f(x) & = & \frac{1}{2a_0}\left[1 + \left(1-\frac{{{\omega}_p^{\prime}}^2 x}{1-{\Omega}^{\prime}x}\right)^{1/2} \right] \times  \nonumber \\
 & &
\left[\frac{\sqrt{1 - x^2}}{x} \left(\Omega^{\prime} x - 1 \right) \right] - 1
\label{non}
\end{eqnarray}
with $x = \sqrt{1 - u_{\pm}^2}$. This equation is numerically solved for $u_{\pm}$ and the values of $\tilde{E}_{\pm}$ and $\tilde{E}_{z}$ thus obtained are compared with simulation, as discussed in the next section. (
Although the above equation yields multiple values of $u_{\pm}$, the simulation results, as expected,  correspond to the lowest positive value. ) 

The analysis becomes comparatively simpler in the non-relativistic limit ($u_{\pm}\ll 1$, $x \approx 1$), which gives
\beea
\tilde u{\pm}= \frac{2 a_0}{(N+1)(\tilde{\Omega}-1)}
\enea
with $a_0=eE_0/m \omega c$.\cite{luan}. In the non-relativistic limit, the refractive index N becomes constant as evident from  Eq. (\ref{n10}). Thus using $\tilde{E}_{\pm} = u_{\pm}(\Omega^{\prime} - 1)$ and $\tilde{B}_{\pm} = \tilde{E}_{\pm} / \beta$, equation (\ref{n8}) gives the expression for electrostatic field, under non-relativistic approximation, as 
\beea
\tilde E_z=-\frac{8 N a_0^2}{(N+1)^2(\tilde{\Omega}-1)}\sin (2k_0 N z)
\label{m1}
\enea
This is exactly the same expression for the static electric field generated by superposition of whistler waves as obtained in the non-relativistic theory developed by Sano {\it et}. al. \cite{sano}. In the next section, we present a comparison of the relativistic theory with 1-D Particle-in-cell simulation results.

\section{Fully Relativistic PIC Simulation: OSIRIS}
Here we present the results obtained using fully relativistic 1D PIC simulation which is performed to study the interaction of two  oppositely moving right circularly polarized electromagnetic waves in a dense plasma. The code used is fully electromagnetic and massively parallel code ``OSIRIS 4.0".\cite{fon,hem} The geometry of the problem is similar to simulation performed by Sano {\it et al.}\cite{sano} Here a slab of preformed plasma of density $\sim 3.37 \times 10^{22}$ /cc corresponding to the solid hydrogen density has been considered. The system is  extended  $\sim 69.5 \,\, \mu m$ along the longitudinal direction. Number of particles per cell is $\sim 200$ and the chosen time step is $\sim 0.005 \,\, \omega_p^{-1}$.  The laser profile is considered to be Gaussian with both longitudinal rise and fall to be of $ \sim 9 \,\, \mu m$ and it is used throughout all our simulation runs. Number of grids are $\sim 3000$ along the longitudinal direction with the resolution of order $\sim \lambda_0/40$. The open boundary conditions are taken for both the fields and the particles. The EM waves are right circularly polarized with the transverse field expressed as 
${\tilde E}_{\perp}=\sqrt[•]{{\tilde E}_x^2+{\tilde E}_y^2}$. Before we discuss the standing wave formation by superposition of counter propagating whistler waves, the transmission characteristics of whistler waves in overdensed plasma medium is investigated first.

\subsection{Transmission of Whistler Waves}

\begin{figure}
{\centering
{\includegraphics[width=3.0in,height=2.2in]{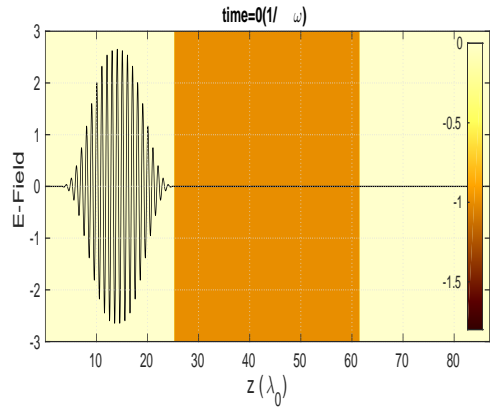}\par}}
\caption{ Incident right circularly polarized EM wave field ($eE_y/m\omega c$) in vacuum corresponding to the parameters: $a_0=2.65$,  $B_0/B_c=7.47, n_{0}/n_c=19.3$.}
\label{fig.1}
\end{figure}

\begin{figure}
{\centering
{\includegraphics[width=3.0in,height=2.2in]{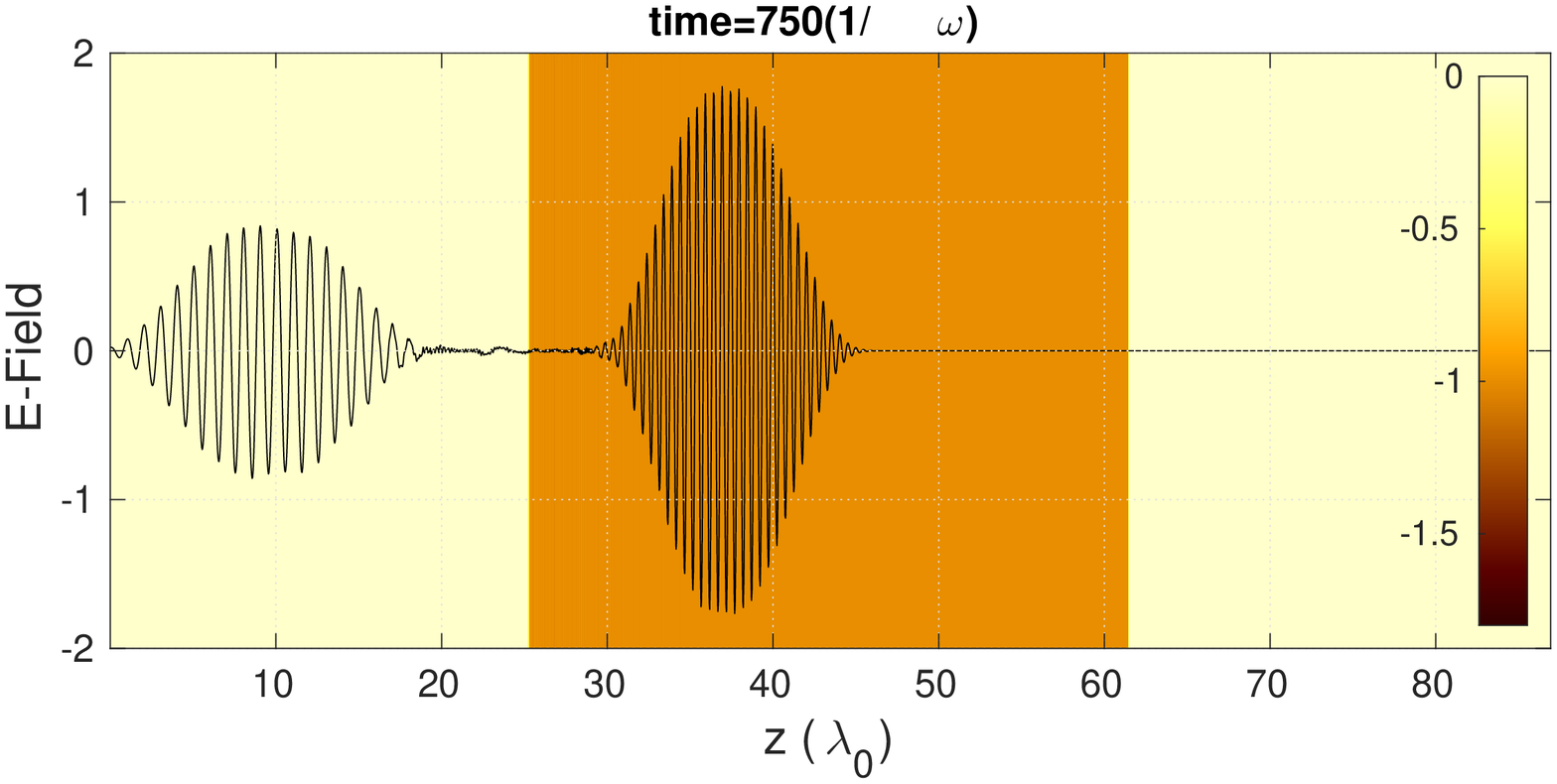}\par}}
\caption{The transmitted and reflected EM wave fields ($eE_y/m\omega c$) in the dense plasma medium corresponding to the parameters $a_0=2.65$,  $B_0/B_c=7.47, n_{0}/n_c=19.3$.}
\label{fig.2}
\end{figure}

\begin{figure}
{\centering
{\includegraphics[width=3.0in,height=2.2in]{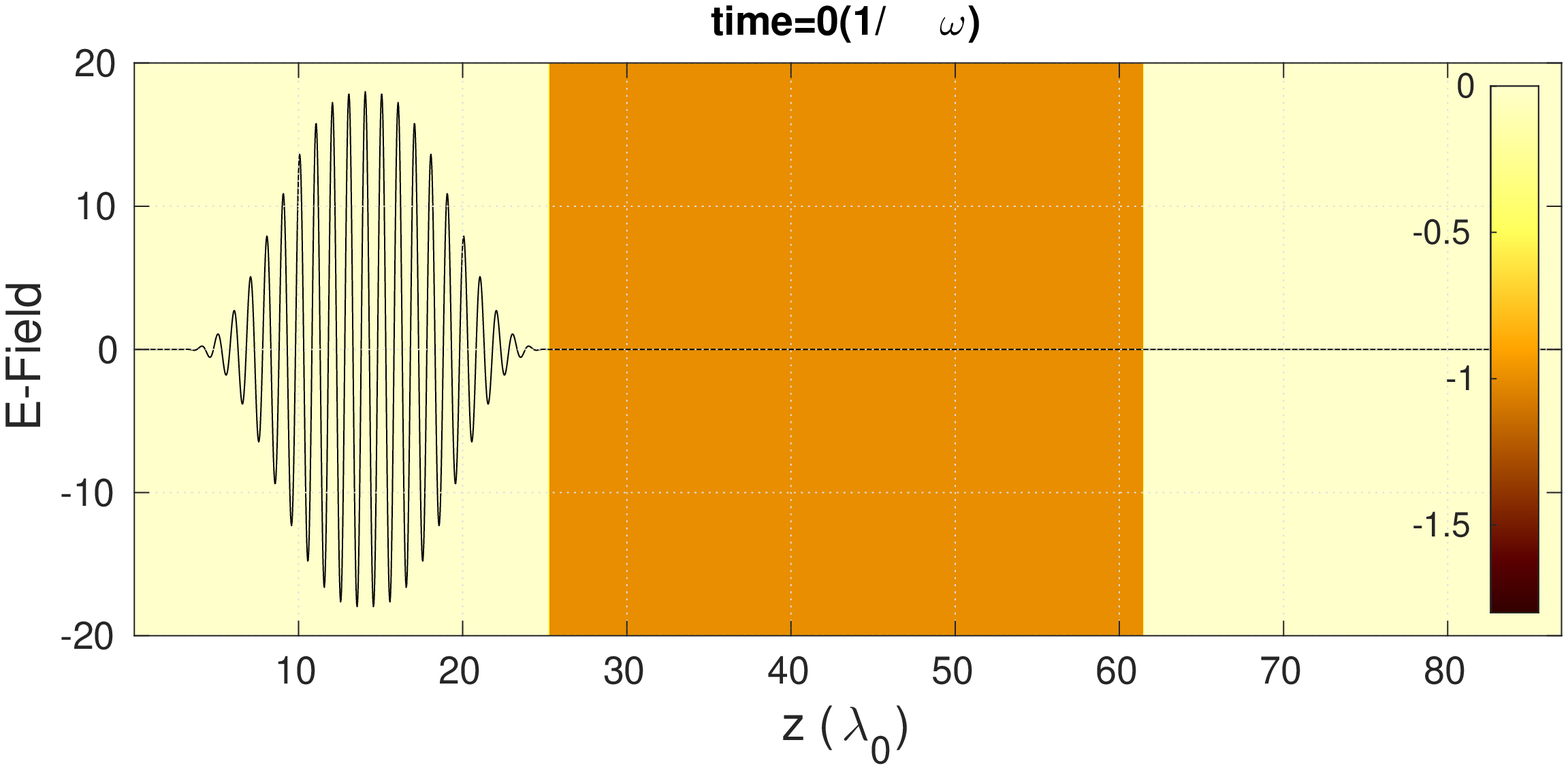}\par}}
\caption{Incident right circularly polarized EM wave field ($eE_y/m\omega c$) in vacuum corresponding to the parameters: $a_0=18.0$,  $B_0/B_c=50.0, n_{0}/n_c=19.3$.}
\label{fig.3}
\end{figure}

\begin{figure}
{\centering
{\includegraphics[width=3.0in,height=2.2in]{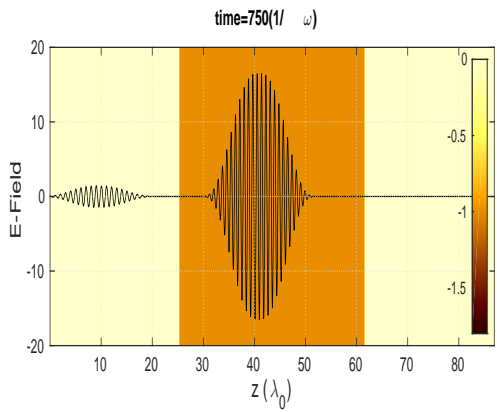}\par}}
\caption{The transmitted and reflected EM wave fields ($eE_y/m\omega c$) in the dense plasma medium corresponding to the parameters: $a_0=18.0$,  $B_0/B_c=50.0, n_{0}/n_c=19.3$.}
\label{fig.4}
\end{figure}

In the first simulation experiment, a single circularly polarized electromagnetic wave is launched from the left side of the plasma slab. The transmitted field amplitude of the wave inside the plasma is determined from the Fresnel equations and the relativistic theory of whistler wave propagation, as discussed in the previous section. The 1D PIC simulation shows that for the chosen parameters $a_0=2.65$,  $B_0/B_c=7.47, n_{0}/n_c=19.3$ with $B_c=m\omega c/e$ and $n_c=(m\omega^2/4\pi e^2)$, the measured maximum field amplitude of the transmitted wave ($eE_y/m\omega c$) is around $\sim 1.7678$. According to the relativistic theory, the calculated transmitted field amplitude is  also around $\sim $ 1.7676 corresponding to the lowest root of fluid velocity ($u=0.20$), thus closely matching the measured value. The field profiles in the vacuum and in the plasma slab are respectively shown in Fig. (\ref{fig.1}) and Fig. (\ref{fig.2}). 

The simulation has also been performed for another set of parameters. The measured maximum transmitted field amplitude for the parameters $a_0=18.0$,  $B_0/B_c=50.0, \,\, n_{0}/n_c=19.3$ is around 
$\sim 16.5074$ which also closely matches with the  relativistic theory ($eE_y/m\omega c \sim 16.5072$) corresponding to the lowest root ($u=0.34$). The results are shown in Fig. (\ref{fig.3}) and Fig. (\ref{fig.4}). In the following subsection we discuss the standing wave formation when we launch right circularly polarized EM waves from both sides of the plasma slab.

\subsection{Superposition of Whistler Waves}

In the next simulation run we consider that the plasma slab is irradiated by two counter propagating right circularly polarized waves launched from both sides in the longitudinal direction. 
The circularly polarized waves are propagating in the 
$\pm z$ direction  after they are launched from the two opposite boundaries.  The electric field of the standing wave generated in the longitudinal direction 
(${\tilde E}_z$)  as a result of the superposition of the two counter propagating EM waves is observed. 

\begin{figure}
{\centering
{\includegraphics[width=3.0in,height=2.2in]{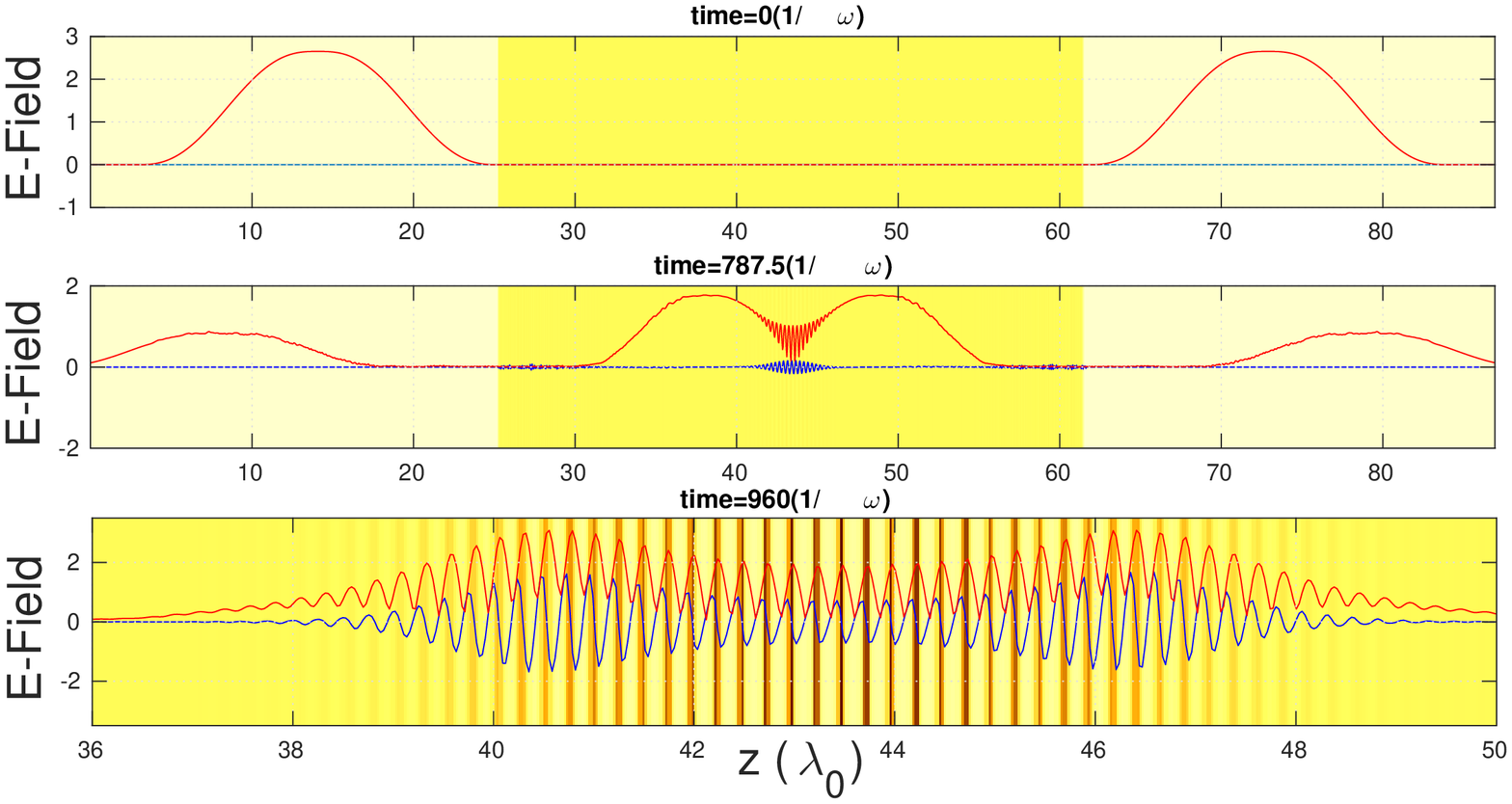}\par}}
\caption{ Snap shots of the longitudinal electrostatic field  $eE_z/m\omega c$ (in blue) and laser electric field $eE_y/m\omega c$ (in red)  from simulation at different time steps. The first panel shows the laser profile 
when the pulses launched from both sides enters into the plasma. The middle and the third panel depicts the starting and formation of the standing wave in the interaction of counter propagating whistler waves.  
The chosen parameters for this simulation run are: $a_0=2.65$,  $B_0/B_c=7.47, n_{0}/n_c=19.3$ with $B_c=m\omega c/e$ and $n_c=(m\omega^2/4\pi e^2$).}
\label{fig.5}
\end{figure}

\begin{figure}
{\centering
{\includegraphics[width=3.0in,height=2.2in]{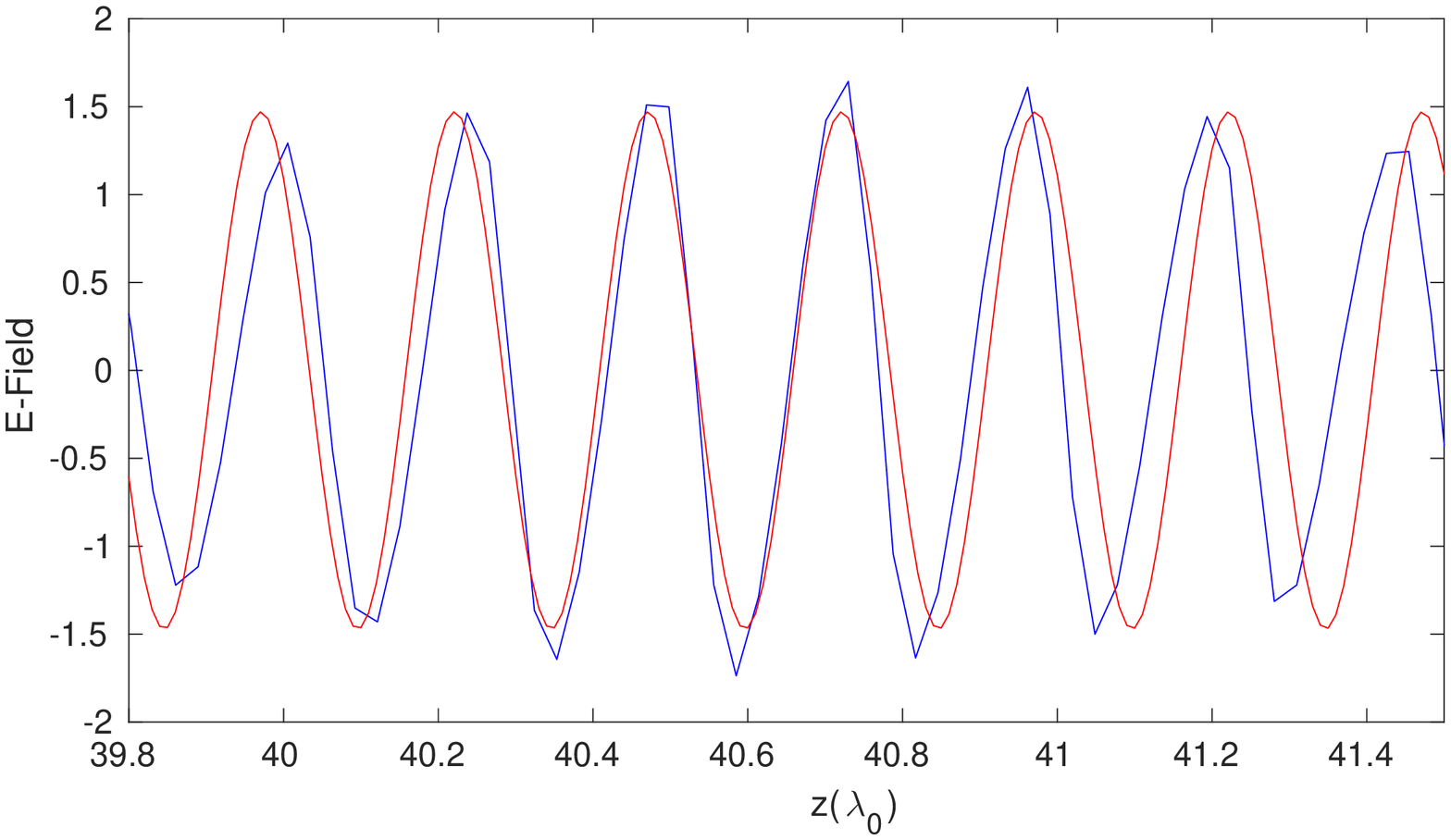}\par}}
\caption{Comparison of the standing wave field profiles obtained from theory with the PIC simulation. The longitudinal electrostatic field ( $eE_z/m\omega c$ ) of the standing wave is represented by blue curve  as obtained from simulation at time 960 $\omega^{-1}$ and by red curve obtained from relativistic theory with the lowest root of fluid velocity ($u=0.20$). The laser amplitude, magnetic field and plasma density  are chosen as $a_0=2.65$,  $B_0/B_c=7.47, n_{0}/n_c=19.3$ with $B_c=m\omega c/e$ and $n_c=(m\omega^2/4\pi e^2$).}
\label{fig.6}
\end{figure}

\begin{figure}
{\centering
{\includegraphics[width=3.0in,height=2.2in]{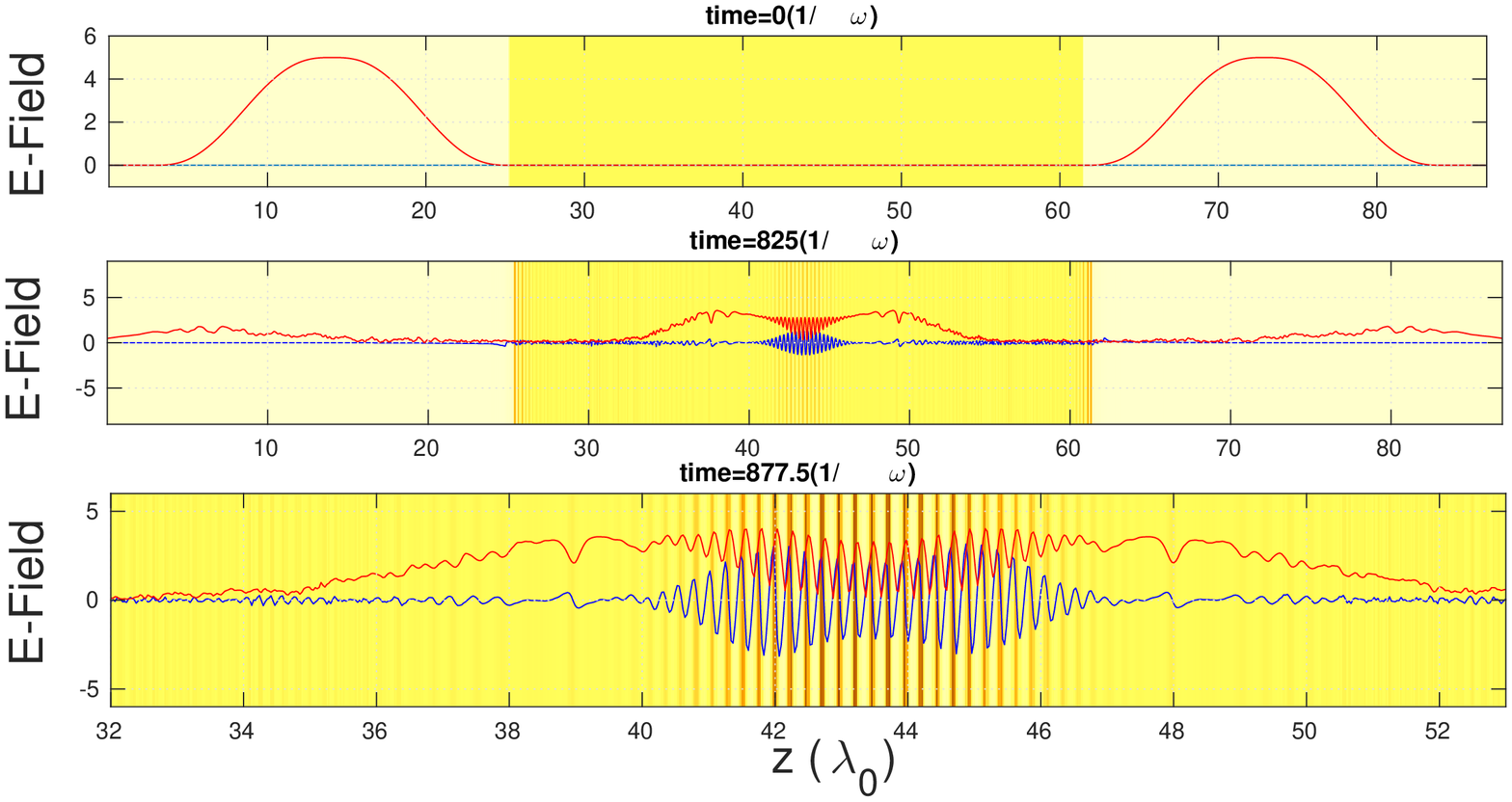}\par}}
\caption{Snap shots of the longitudinal electrostatic field  $eE_z/m\omega c$ (in blue) and laser electric field $eE_y/m\omega c$ (in red)  from simulation at different time steps.
We choose $a_0=5.0$,  $B_0/B_c=7.47, n_{0}/n_c=19.3$.}
\label{fig.7}
\end{figure}

\begin{figure}
{\centering
{\includegraphics[width=3.0in,height=2.2in]{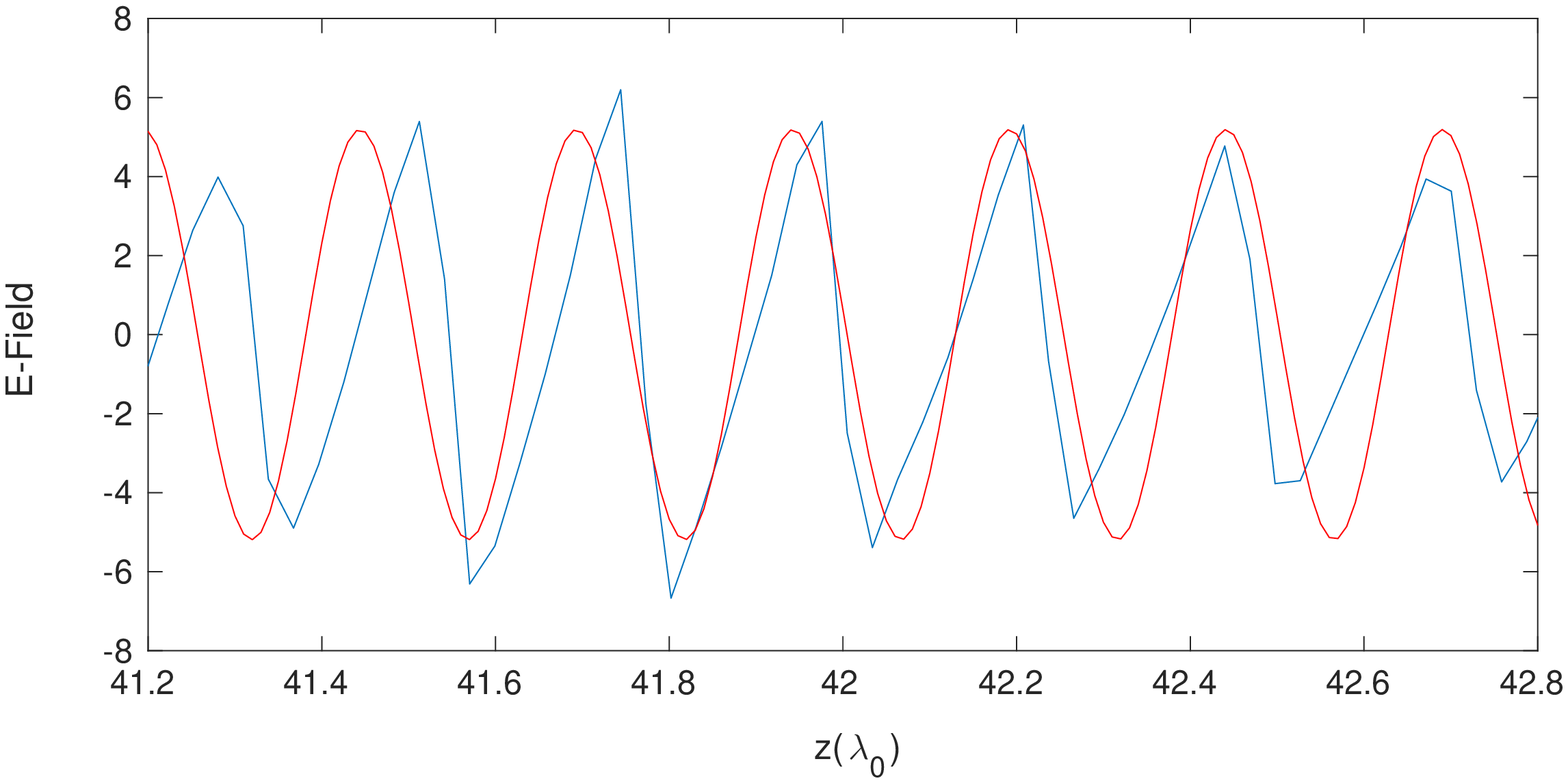}\par}}
\caption{Longitudinal electrostatic field  $eE_z/m\omega c$  from simulation (in blue) at 
time 877.5 $\omega^{-1}$ and  from relativistic theory with root $u=0.38$ (in red).
Here we choose $a_0=5.0$,  $B_0/B_c=7.47, n_{0}/n_c=19.3$.}
\label{fig.8}
\end{figure}

In this simulation experiment we consider an external magnetic field of strength $\sim 100\,\,\,kT$ which is applied along the propagation direction. A laser pulse with relativistic intensity of $a_0=2.65$ is taken, corresponding to a laser  intensity of $I=3\times10^{19}$ W/cm$^2$ with 30 fs pulse-width and wavelength of 0.8 $\mu$m. The simulation results are shown in Fig. (\ref{fig.5} ) and Fig (\ref{fig.6}). It is observed that the normalized maximum electrostatic field amplitude of the excited standing wave is around $\sim 1.5$ for $a_0=2.65$ as evident from Fig. (\ref{fig.5}). 
The slow spatial variation of electrostatic field $E_z$ seen in simulation is due to the Gaussian envelope of the incoming light pulse.
This maximum amplitude matches with the field amplitude as predicted by the relativistic theory [Fig. (\ref{fig.6})] corresponding to the lowest root $u_{\pm}=0.20$ obtained by solving Eq. (\ref{non}).  
This differs from the maximum field amplitude obtained using non-relativistic theory which predicts $e E_z/m\omega c \sim 1.9$. This disparity occurs because in the non-relativistic situation the refractive index $N$ is constant and related to the electric field amplitude via a simple relation $E=2a_0/(N+1)$. 
%
%
Therefore, it should be emphasized that the relativistic theory which is developed without any approximation is more accurate and explains the correct physics of standing wave formation. This assertion is confirmed by our 1D PIC simulation results.

\begin{figure}
{\centering
{\includegraphics[width=3.0in,height=2.2in]{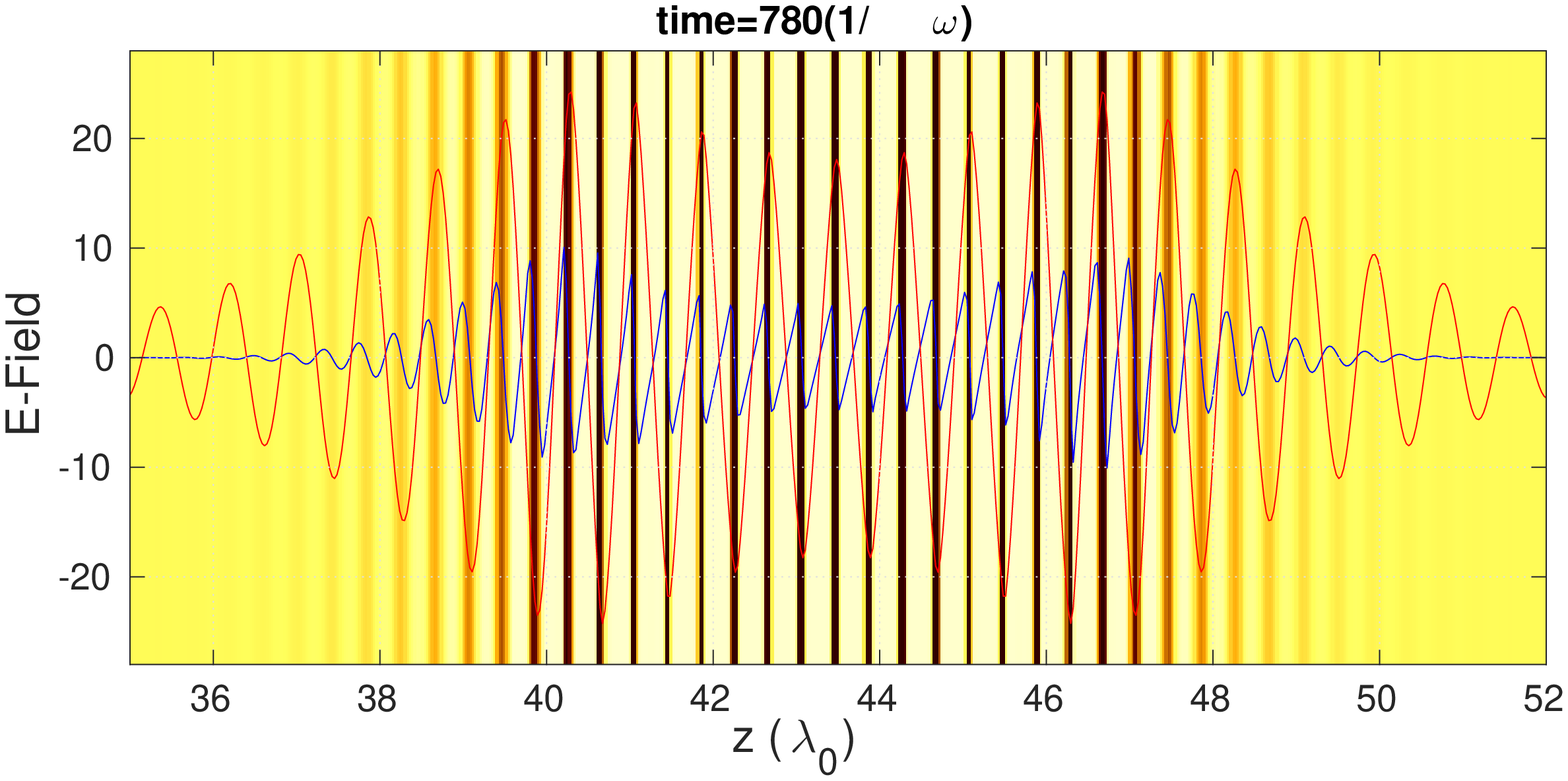}\par}}
\caption{Longitudinal electrostatic field  $eE_z/m\omega c$ (in blue) and laser electric field (in red) from simulation .
Here we choose $a_0=15.0$,  $B_0/B_c=45.0, n_{0}/n_c=19.3$.}
\label{fig.9}
\end{figure}

\begin{figure}
{\centering
{\includegraphics[width=3.0in,height=2.2in]{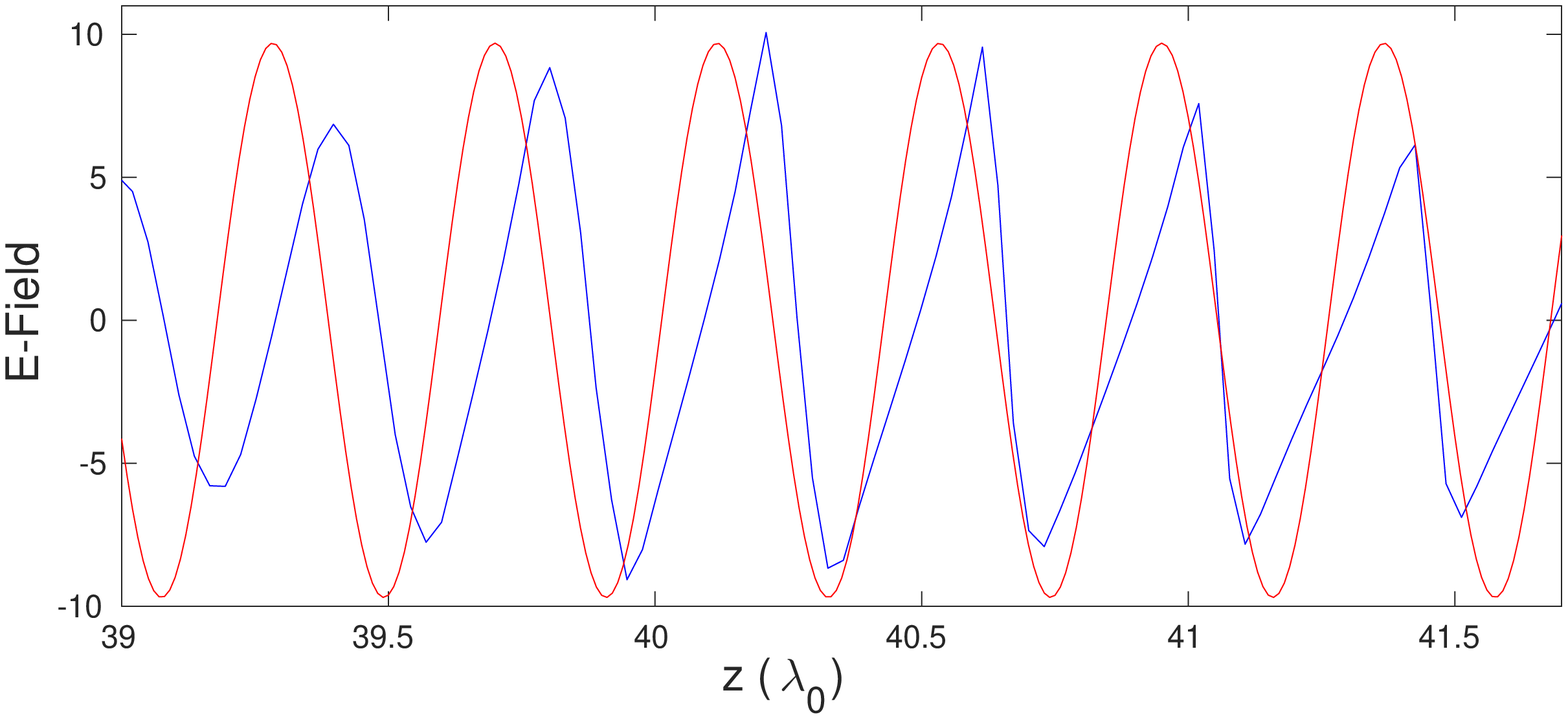}\par}}
\caption{Comparison of longitudinal electrostatic field  $eE_z/m\omega c$  from simulation (in blue) at time 780 $\omega^{-1}$ and  from relativistic theory with lowest root $u=0.32$ (in red).
The chosen parameters are $a_0=15.0$,  $B_0/B_c=45.0, n_{0}/n_c=19.3$.}
\label{fig.10}
\end{figure}

\begin{figure}
{\centering
{\includegraphics[width=3.0in,height=2.2in]{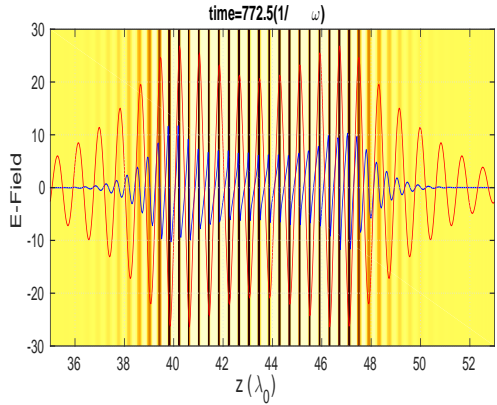}\par}}
\caption{Longitudinal electrostatic field  $eE_z/m\omega c$ (in blue) and laser electric field (in red) from PIC simulation. Here we choose $a_0=18.0$,  $B_0/B_c=50, n_{0}/n_c=19.3$.}
\label{fig.11}
\end{figure}

\begin{figure}
{\centering
{\includegraphics[width=3.0in,height=2.2in]{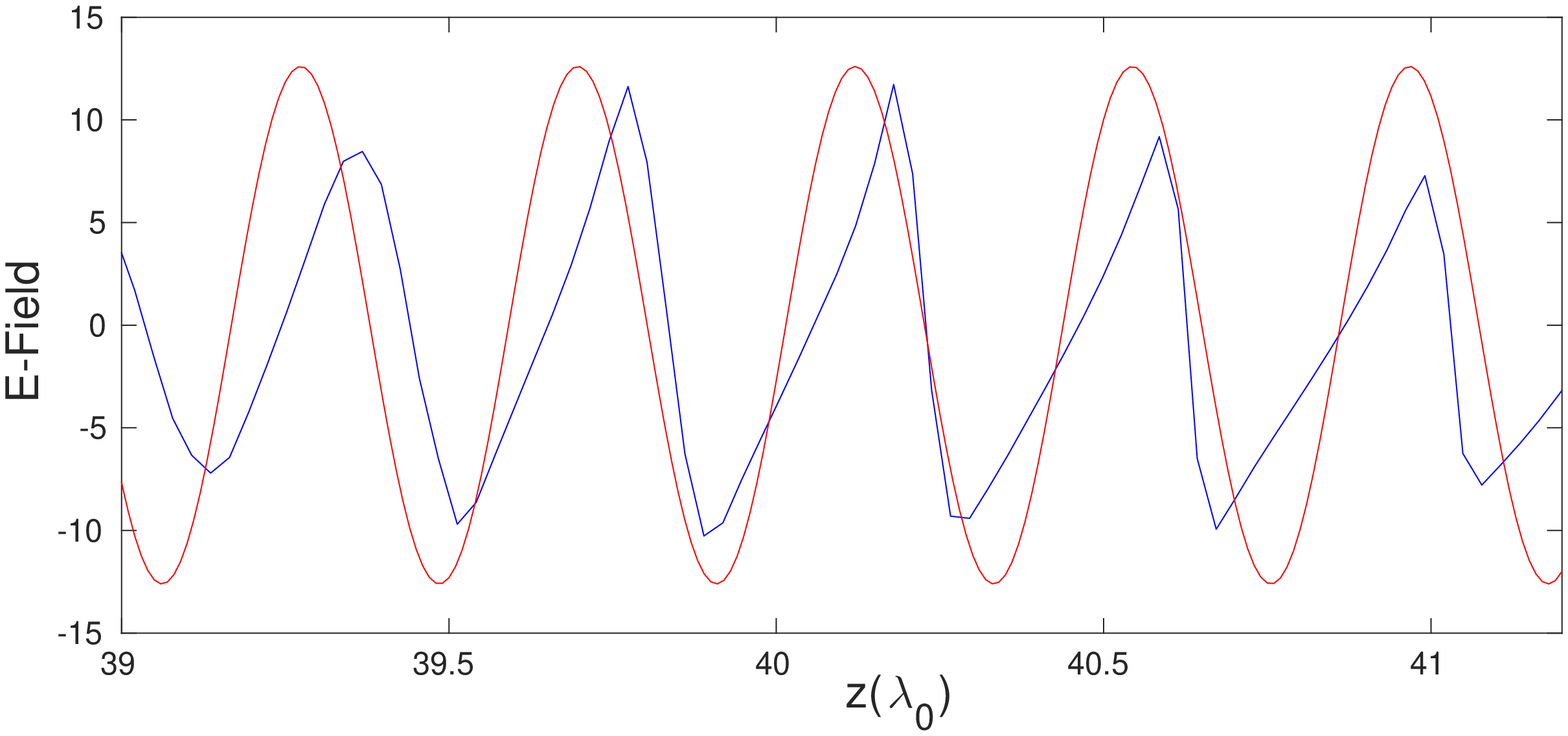}\par}}
\caption{Comparison of longitudinal electrostatic field  $eE_z/m\omega c$  from simulation (in blue) at time 772.5 $\omega^{-1}$ and  from relativistic theory with lowest root $u=0.34$ (in red).
Here we choose $a_0=18.0$,  $B_0/B_c=50.0, n_{0}/n_c=19.3$.}
\label{fig.12}
\end{figure}

The simulation has also been performed with three other sets of parameters. For the parameter set ($a_0=5.0$,  $B_0/B_c=7.47, n_{0}/n_c=19.3$), the lowest root obtained from relativistic theory is $u_{\pm} = 0.38$.
%
%
In this case the laser amplitude is $1.1\times 10^{20}$ W/cm$^2$. Similar to the previous case the field amplitude corresponding to this lowest root gives the closest matching with the simulation result [Fig. (\ref{fig.7}) and (\ref{fig.8})]. Same conclusion can be drawn for the other set ($a_0=15.0$,  $B_0/B_c=45, n_{0}/n_c=19.3$) as well, which is evident from the Fig. (\ref{fig.9}) and
Fig. (\ref{fig.10}). The chosen parameters here correspond to the magnetic field of $\sim 602 \,\, kT$ with laser amplitude of $9.7\times 10^{20}$ W/cm$^2$. In addition, the results corresponding to magnetic field of $\sim 670 kT$ and laser amplitude of $1.4\times 10^{21}$ W/cm$^2$ [$a_0=18.0$,  $B_0/B_c=50, n_{0}/n_c=19.3$] is also shown in  Fig. (\ref{fig.11}) and Fig. (\ref{fig.12}). In all cases, simulation results closely match with the relativistic theory of electrostatic standing wave formation.

\section{conclusion}
We have presented a general theory of standing wave excitation in the superposition of 
two counter propagating relativistic whistler waves. This theory is supplemented by fully relativistic electromagnetic 1D PIC simulations performed using the code OSIRIS 4.0. In the limiting 
cases we recover  results established by Sano {\it et al.}\cite{sano} in the non-relativistic limit. 
However, in contrast to the non-relativistic theory, it is found that the relativistic theory predicts a lower value of electrostatic field
for a particular set of parameters.
It must be emphasized that correct explanation of the interaction physics can be obtained only with the relativistic theory. 
The occurrence of lower value of electrostatic field may be understood as a 
consequence of relativistic increase of electron mass making them ``sluggish'', thus resulting in lower values of electron current. This leads to lower amplitudes of Whistler fields, which in turn reduces the amplitude of the electrostatic fields. Further studies related to standing wave breaking and consequent ion heating will be presented in a future publication.

\section{ACKNOWLEDGMENTS}
The authors are grateful to the IPR computer center where the
simulation studies presented in this paper were carried out. We would like to acknowledge the OSIRIS Consortium, consisting of UCLA and IST (Lisbon, Portugal), for providing access to the
OSIRIS 4.0 framework. \cite{fon2,fon3}



\end{document}